\documentclass[12pt,preprint]{emulateapj}

\usepackage{graphicx}
\usepackage{color}

\newcommand{\beq}{\begin{equation}}
\newcommand{\eeq}{\end{equation}}
\newcommand{\beqa}{\begin{eqnarray}}
\newcommand{\eeqa}{\end{eqnarray}}

\begin{document}

\title{\bf  Anomalous bremsstrahlung and the structure of cosmic ray electron-positron fluxes at the GeV-TeV energy
range\\
}

\author{Wei Zhu$^1$\altaffilmark{*}, Peng Liu$^1$, Jianhong Ruan$^1$, Lei Feng$^2$ and Fan Wang$^3$}

\affiliation{$^1$Department of Physics, East China Norma University,
Shanghai 200241, China\\
$^2$ Key Laboratory of Dark Matter and Space
Astronomy,\\
Purple Mountain Observatory, Chinese Academy of Sciences, Nanjing
210008, China\\
Colege of Physics, Qingdao University, Qingdao 266071, China\\
$^3$Department of Physics, Nanjing University, Nanjing,210093, China
} \altaffiltext{*}{ Corresponding author: wzhu@phy.ecnu.edu.cn}

\begin{abstract}

     We reveal that the energy spectra of electrons-positrons in primary
cosmic rays measured at atmosphere top have double structures: an
excess component $\Phi^s_{e^+}(E)=\Phi^s_{e^-}(E)$ around $400~GeV$,
which origins from a strong $e^+e^-$-source and the distorted
background $\Phi^0_{e^-}(E)$.  We supposed that the difference between
AMS-CALET and Fermi-LAT-DAMPE data origins from the energy loss of
the fluxes due to the anomalous bremsstrahlung effect at a special
window. The evolution of spectra under anomalous bremsstrahlung
effect satisfies an improved electromagnetic cascade equation. The
above spectra are parameterized and they can be regarded as the
subjects exploring new physics. We suggest to check the previous
applications of the Bethe-Heitler formula in the study of the
propagation of high energy electrons and photons.

\end{abstract}


\setlength{\parindent}{.25in}

\section{Introduction}

 The precise measurements of cosmic ray electron-positron fluxes by
Alpha Magnetic Spectrometer (AMS) (Aguilar et al. 2013;Aguilar et
al. 2018;Aguilar et al. 2019b;Aguilar et al. 2019a), Fermi Large
Area Telescope (Fermi-LAT) (Abdollahi 2017), DArk Matter Particle
Explorer (DAMPE) (Ambrosi et al. 2017) and Calorimetric Electron
Telescope (CALET) (Adriani 2018) present a complex structure of the
energy spectra at the GeV-TeV energy range. In particular, AMS
reports that the positron flux has a significant excess peaked at
$\sim 300~ GeV$ (or $\sim 400~GeV$ if subtracting the contribution
of the diffuse background), while the electron flux exhibits a
single power law on the same energy range. On the other hand, DAMPE
observed a break of the total electron and positron spectrum at
$\sim 700~GeV$. A confusing problem is that the data of AMS and
CALET are noticeably lower than that of DAMPE and Fermi-LAT in the
above mentioned energy band (Fig. 1). Obviously, if the cause of
this difference is not clarified, any research based on these data
will lose their credibilities.

    Both AMS and CALET set on the international
space station at $\sim 400~km$ height, while Fermi-LAT and DAMPE are
orbiting the Earth at $500\sim560~km$ altitude. Charged particles
may interact with atoms in atmosphere and form the electromagnetic
shower. In the normal condition the radiation length $\lambda\simeq
37~g/cm^2$, it corresponds to a real distance
$\Lambda=\lambda/\rho_c\simeq 4\times 10^{15}cm$ since the
atmospheric density $\rho_c\sim 10^{-14}g/cm^3$ at $400\sim 500~km$
altitude (Qin et al. 2004, Zheng et al. 2007). The altitude
difference $100~km$ is only $1/100000000\times\Lambda$. Therefore,
according to the traditional electromagnetic radiation theory, the
difference of the above energy spectra can't be attributed to the
electromagnetic shower.

\begin{figure}
    \begin{center}

        \includegraphics[width=0.5\textwidth]{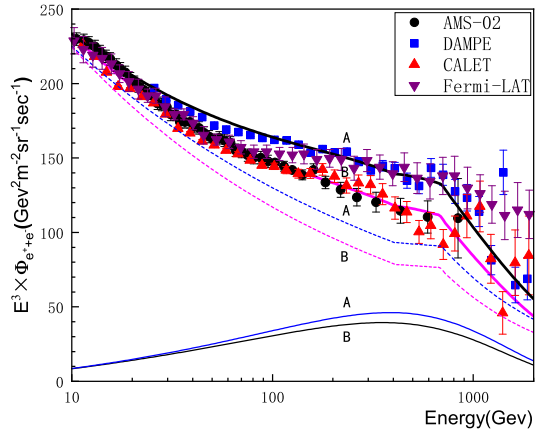}
        \caption{Cosmic electron-positron spectra multiplied by $E^3$ as a
function of energy. A and B indicate the spectra at height $\sim
500$ and $\sim 400$ km. The thin solid curves are the source fluxes,
the dashed curves are the background and the thick solid curves are
their sum. All curves from A to B satisfy the improved cascade
equation with $0.13$ radiation length}\label{fig:1}

    \end{center}
\end{figure}

    One of the theoretical basis of electromagnetic shower is bremsstrahlung.
When charged particles scatter off electric field of proton or
nucleus, they can emit real photons. This is bremsstrahlung (braking
radiation). A quantum-mechanical description of the bremsstrahlung
emission is the famous Bethe-Heitler (BH) formula, which is proposed
in 1934 (Bethe et al. 1934) and it has been widely applied in many
branches of physics and astrophysics. In a previous work (Zhu 2020),
one of us (WZ) pointed out that the BH formula should be modified to
a new form in thin ionosphere, where the integrated bremsstrahlung
cross section has an unexpected big increment, which is large enough
to cause significant reduction of the electron and positron fluxes
in the range of $100~km$. Based on this discovery, an improved
cascade equation for electromagnetical shower was established in
(Zhu 2020).

    Bremsstrahlung dominates the energy loss of high energy electrons in
electromagnetic cascade shower. Bremsstrahlung occurs in the
electric field of a nucleus, the processes will be screened by the
atomic electrons for impact parameters larger than the radius of the
atom (or the Debye length) for neutral atom (or ionized atom).

    The condition applied the anomalous bremsstrahlung formula is that
target recoil can be neglected for a very high energy incident
electron (Zhu 2020). In this case, $1/m^2_e-$dependent integrated
cross section of the BH formula is replaced by $1/\mu^2-$dependence,
$1/\mu$ is the electromagnetic screening scale. This is the
anomalous bremsstrahlung effect. In other environments,
bremsstrahlung still obeys the normal BH formula due to recoil of
target.

    A completely ionized thin gas with a lower density is an ideal place
for testing the anomalous bremsstrahlung effect since ionized atoms
have larger Debye length. The nuclear Coulomb potential may spread
into such a broader space, where the bremsstrahlung events can
neglect the recoil since the incident high energy electrons are far
away from the nucleus. Bremsstrahlung in this environment exhibits a
big cross section since $\mu\ll m_e$. However, the bremsstrahlung
probability also decreases with reduced density and it disappears in
the extremely thin ionosphere. The above two opposite effects lead
to a limited density-window, where the bremsstrahlung process
becomes anomalous.

    According to the data in Fig. 1 we suggest that atmospheric
ionosphere around $400\sim 600~km$ belongs to this observing window,
where the density of complete ionized oxygen is about $\rho_c\sim
10^{-14}g/cm^3$. It meas that the BH formula is valid at the most of
the atmosphere where $\rho\gg \rho_c$. On the other hand, the
bremsstrahlung effect is not important for a very thin gas (for say
$\rho\ll \rho_c$) even the gas is ionized. Further more, we will
show that when electrons traveling from 500 $km$ height to 400 $km$
height, the flight distance is much smaller than the radiative
length. It implies that the incident electrons with high energy may
easily pass through the above anomalous window.

    Note that most of the atmosphere at much lower than 400 $km$ is made
up of neutral atoms, or ionized atoms (Grupen 2005) but they do not
have sufficiently large Debye length since $\rho\gg \rho_c$. When
$GeV\sim TeV$-electrons pass through these atoms and radiate
photons, the nuclear recoil is non-negligible since a smaller impact
parameter. Therefore, the experiments based on the ground, which
observe the electromagnetic shower development are insensitive to
the anomalous bremsstrahlung effect. A best way is to directly
compare the electron spectra measured at two different heights in
ionosphere around $\rho_c\sim 10^{-14}g/cm^3$ as shown in Fig.1.

    In this letter we analyze the measured data of
electron-positron fluxes. We find that there is a same excess source
in both electron- and positron-fluxes, which is peaked at $\sim
400~GeV$. On the other hand, the background of electrons has the
broken power form: from $10~GeV$ to $400~GeV$ the spectrum obeys a
single power law, after $400~GeV$ the spectrum is hardened somewhat,
then it again becomes steep (see the dashed curves in Fig. 1). The
evolutions of the above fluxes from set A to set B satisfy
quantitatively the prediction of the improved cascade equation (Zhu
2020). We will give our arguments in Section 2. The discussions and
summary are given in Section 3

\section{Reconstruction of the experimental data}

      We divide the measured electron-positron spectra $\Phi_{e^++e^-}$ in Fig. 1
into two sets with different observation heights: A (DAMPE-Fermi-LAT
at $\sim$ 500 km height and B (AMS-CALET at $\sim$ 400 km height).
We begin from the AMS data about positron spectrum. The background
spectrum $\Phi^{0,B}_{e^+}(E)$ of positrons is described at the
lower energy range in the diffuse model. Therefore, we can extract
the background $\Phi^{0,B}_{e^-}(E)$ (or $\Phi^{0,B}_{e^++e^-}(E)$)
using the difference between the measured spectra
$\Phi^{B}_{e^++e^-}(E)$ and $\Phi^{s,B}_{e^+}(E)$.  Where "0" and
"s" mark diffusion background and astrophysical no-diffusion sources
like dark matter or others.

    We define the source as the predictions of dark
matter model or gluon condensation (GC) model, where
$\Phi^s_{e^+}=\Phi^s_{e^-}$; all remaining components are regarded
as the background $\Phi^0_{e^+}$ and $\Phi^0_{e^-}$.

    AMS has parameterized the positron flux
of the source as (Aguilar 2019a)

$$\Phi^{s,B}_{e^+}(E)=\frac{E^2}{\hat{E}^2}6.8\times 10^{-5}(\hat{E}/60)^{-2.58}exp(-\hat{E}/813),\eqno(2.1)$$
where $\hat{E}=E+1.1 GeV$. Its shape is described by the thin curve
B in Fig. 1. The background $\Phi^{0,B}_{e^++e^-}(E)$ is obtained by
subtracting $2\Phi^{s,B}_{e^+}(E)$ from the data of
$\Phi^B_{e^++e^-}(E)$. We find that $\Phi^{0,B}_{e^++e^-}(E)$
presents a single power law $\sim E^{-0.283}$ till $E\sim
300-400~GeV$, then it deviates from this form at more higher energy.
However, the large errors in the data of $\Phi^B_{e^++e^-}(E)$ at
$E>400~GeV$ hinder the extraction of the information. We noticed
that the DAMPE data for $\Phi^A_{e^++e^-}(E)$ have a break around
$E\sim 700~GeV$. If set A and set B are connected by the
electromagnetic cascade, the flux $\Phi^{0,B}_{e^++e^-}(E)$ will
roughly keep the shape in a short cascade length ($t\ll 1$).
Therefore, we have the following form

$$\Phi^{0,B}_{e^++e^-}(E)=\left\{
\begin{array}{ll}
431E^{-0.283}
& {E<400~GeV}\\
108E^{-0.053} & {400~GeV\leq E\leq 700~GeV}\\
12500E^{-0.783} & {E>700~GeV}
\end{array}\right..\eqno(2.2)$$ One can find that
$\Phi^B_{e^++e^-}(E)= 2\Phi^{s,B}_{e^+}(E)+\Phi^{0,B}_{e^++e^-}(E)$
(the thick curve in Fig. 1) is consistent with the data. This result
shows that $\Phi^B_{e^++e^-}(e)$ has a double structure: a new
source creating high-energy electron-positron pair and the electron
background, the later is distorted in propagation.

    Now we derive the structure of electron and positron fluxes at $\sim 500~km$
using the improved cascade equation. The elemental high-energy
processes that make up electromagnetic cascade are pair production
and bremsstrahlung (Gaisser et al. 2016). We concern the energy
spectra of electrons and positrons at high energy ($E>10~GeV$) in
this work. Therefore, the contributions of $\gamma\rightarrow
e^++e^-$ can be neglected. We denote $X$ and $\lambda$ as the depth
and the radiation length in unity $g/cm^2$. The contributions of
$\gamma\rightarrow e^++e^-$ to the electromagnetic showers at the
high energy band can be neglected.  Omitting the upper indexes, the
fluxes of electron-positron in (2.1) and (2.2) are denoted by
$\Phi_e$ ($e=e^+,e^-$) and they satisfy the following cascade
equation (Zhu 2020)

$$\frac{d\Phi_e(E,
t)}{dt}=\int^{\infty}_{E} \frac {dE'}{E'}P_{e\rightarrow
e}\left(\frac{E}{E'}\right)\Phi_e(E',t)$$
$$- \Phi_e(E,t)\int_0^1
dzP_{e\rightarrow e}(z).\eqno(2.3)$$ Where $P_{e\rightarrow
e}(z)dtdz$ is the probability for an electron/positron of energy
$E'$ reduces to energy $zE'$ after radiation in traversing
$dt=dX/\lambda$ (Zhu 2020),

$$P_{e\rightarrow e}(z)=\frac{3}{4}\frac{1+z^2}{1-z^2}. \eqno(2.4)$$

     Because we take the radiation length
$\lambda=Am_p/\sigma$ ($Am_p$ is atom mass) as the unit of cascade
step, equation (2.3) satisfies the electromagnetical showers either
with or without the anomalous bremsstrahlung effect.
In the calculation, we take the spectra A as the input at the
starting point $t=0$. We find that spectra A evolve to $B$ at
$t=0.13$ in Fig. 1.  Note that the anomalous bremsstrahlung
effect happens at high energy range, while the BH formula is
valid at lower energy, where the anomalous bremsstrahlung effect
disappears. We use a simple function $\chi(E)=0.5\log (E/1~GeV)-0.5$
at $(10~GeV<E<1000~GeV)$ to connect the solutions between low and
high energy ranges, i.e.,

$$\Phi^{s,B}_{e^+}(E,t=0.13)\equiv\Phi^{s,A}_{e^+}(E,t=0)$$
$$-\chi(E)[\Phi^{s,A}_{e^+}(E,t=0)-
\tilde{\Phi}^{s,B}_{e^+}(E,t=0.13)], \eqno(2.5)$$and

$$\Phi^{0,B}_{e^++e^-}(E,t=0.13)\equiv\Phi^{0,A}_{e^++e^-}(E,t=0)$$
$$-\chi(E)[\Phi^{0,A}_{e^++e^-}(E,t=0)-
\tilde{\Phi}^{0,B}_{e^++e^-}(E,t=0.13)], \eqno(2.6)$$where
$\tilde{\Phi}^{s,B}_{e^+}(E,t=0.13)$ and
$\tilde{\Phi}^{0,B}_{e^++e_-}(E,t=0.13)$ are the solutions of the
cascade equation at $t=0.13$. Finally, we find following forms (Fig.
1)

$$\Phi^{s,A}_{e^+}(E,t=0)=\frac{E^2}{\hat{E}^2}7.4\times 10^{-5}(\hat{E}/60)^{-2.54}exp(-\hat{E}/813),\eqno(2.7)$$and

$$\Phi^{0,A}_{e^++e^-}(E,t=0)$$
$$=\left\{
\begin{array}{ll}
389E^{-0.238}
& {E<400~GeV}\\
125E^{-0.050} & {400~GeV\leq E\leq 700~GeV}\\
11369E^{-0.693} & {E>700~GeV}
\end{array}\right..\eqno(2.8)$$

    Now let's explain why the anomalous bremsstrahlung effect
predicts a big difference between the spectra A and B.
$X/\rho_c=100~km$ is the flight distance from A to B. The result
$t=0.13$ means that the cascade from A to B through
$0.13\Lambda\equiv100~km$. Thus, we have $\lambda=\Lambda\rho_c=
10^{-6}g/cm^2$. Obviously, this value is much smaller than the
standard bremsstrahlung result based on the BH formula
$\lambda_{BH}=37 g/cm^2$ (Gaisser et al.2016). Because $\lambda$ is
inversely proportional to the bremsstrahlung cross section
($\lambda=Am_p/\sigma$), it implies that an actual bremsstrahlung
cross section at $400~km\sim 500~km$ is much larger than the
estimation based on the BH formula.

    We consider that this big
difference in the cross section origins from the anomalous
bremsstrahlung effect, i. e., $\lambda_{improvedBH}\equiv
10^{-6}g/cm^2$. A main difference between the BH formula and the
improved BH formula is a factor substitution in their cross sections

$$\frac{1}{m^2_e}\rightarrow \frac{1}{\mu^2}, \eqno(2.9)$$where
$1/\mu\geq 10^{-8}cm\gg 1/m_e$. Therefore, we have
$\lambda_{improvedBH}\ll\lambda_{BH}$.

    Of cause, the above mentioned estimation is only an ideal situation,
where all scattering in an atom has the anomalous bremsstrahlung
effect. Note that the improved BH formula is valid if the recoil is
ignored (Zhu 2020). Therefore, one can imagine that the anomalous
bremsstrahlung effect occurs in the scattering away from the
nucleus. Even so, there is still a sufficiently larger enhanced
cross section to reach $\lambda_{improved-BH}\simeq 10^{-6}g/cm^2$.
Detailed estimation is dependent on the model.

    For comparison, we consider that there is no anomalous bremsstrahlung
effect. As we have mentioned in introduction, according to the
normal bremsstrahlung theory, a high energy electron in a gas with
the density $10^{-6}g/cm^2$ can flight an effective distance
$\Lambda_{BH}=\lambda_{BH}/\rho=10^{10}~km$. The distance $100~km$
from A to B is only $10^{-8}\times\Lambda_{BH}$. Therefore, the
energy loss due to bremsstrahlung is completely negligible and the
data A and B should almost coincide.

There are other inelastic scattering of electrons with
atmosphere, and they can lose energy of electrons.  For example, the
inverse Compton scattering, it needs dense low-energy target photon;
the synchrotron radiation of electrons, which requires a strong
magnetic field. Obviously, the atmosphere cannot provide such dense
photons and strong magnetic field. Besides, the cross sections of
both them are proportional to $\propto 1/m_e^2\ll1/\mu^2$.
Therefore, either the inverse Compton scattering or the synchrotron
radiation cannot obviously hinder the propagation of high-energy
electrons in ionosphere. Thus, we conclude that in our current
knowledge the anomalous bremsstrahlung effect obeying the improved
BH formula is an acceptable explanation of the structure of cosmic ray
electron-positron fluxes in Fig. 1.

\section{Discussions and summary}

    A modified formula $\chi(E)$ can improve the agreement
between the curve $A$ and the experimental data at $E<100~GeV$.
Although $\chi(E)$ is a phenomenological formula, the improved
cascade equation dominates the spectrum structure at the key energy
range $100~GeV<E<1~TeV$. The $\chi^2$ test comparing the described
(or predicted) and observed electron+positron flux is given by

$$\chi^2=\sum_i\frac{(\Phi^{des/pre}_i-\Phi_i^{obs})^2}{\sigma^2_{sys,i}+\sigma^2_{stat,i}},
\eqno(3.1)$$where i runs over the energy range $100~GeV<E<1~TeV$. As
an input we have $\chi^2=3.1$ for the described flux in set B by
AMS-02. We obtain $\chi^2=20.6$ (or $\chi^2/d.o.f.=20.6/16$). Thus,
we recognize that the data of set A and set B reflect a same double
structure. The parameterized formulas (2.7) and (2.8) are closer to
the real signals and they can be regarded as the objects exploring
new physics. We predict that the anomalous effect of bremsstrahlung
and pair creation may arise a big enhancement of the increasing
cross section at the thin ionized gas.

    If the anomalous bremsstrahlung effect is correct, we predict
that the measured energy spectra of cosmic-ray electrons above $\sim
500~km$ or between $400\sim500~km$ altitudes will be higher than the
Fermi-LAT and DAMPE data, or are locked between the Fermi-LAT-DAMPE
and AMS-CALET data. On the other hand, according to this work, there
is a limited density-window for observing the anomalous
bremsstrahlung effect. We suggest that the anomalous bremsstrahlung
effect may appear when high energy electrons and relating
$\gamma$-ray pass through thin ionized gas in supernova remnant
(SNR), active galactic nuclei (AGN) or other relating subjects
(Stickforth 1961,Moskalenko et al. 1997) if where the anomalous
bremsstrahlung effect works. However, what needs to be considered is
how to distinguish this window from a complex ionized gas system. We
will study it later.

   In summary, primary cosmic ray electrons and positrons
enter to the top of atmosphere, their energy spectra may changed by
an anomalous bremsstrahlung effect. This leads to a complex
structure of the measured spectra. Using an improved electromagnetic
cascade equation we present the existence of the anomalous effect in
bremsstrahlung and pair creation, and find that the resulting
spectra appear a double structure. All explorations based on
AMS-CALET-Fermi-LAT-DAMPE data should take into account this double
structure. We also suggest to check the previous applications of the
BH formula in the study of the propagation of electrons and photons.

\noindent {\bf ACKNOWLEDGMENTS}This work is supported by the
National Natural Science of China (No.11851303). FL was supported by
the National Natural Science Foundation of China (No. 11773075) and
the Youth Innovation Promotion Association of Chinese Academy of
Sciences (No. 2016288).

\end{document}